\documentclass[proceedings]{JHEP3}
\usepackage{eurosym}
\usepackage{amssymb}
\usepackage{amsfonts}
\usepackage{amsmath,epsfig}
\usepackage{graphicx}

\newbox\mybox

\newcommand\fverb{\setbox\mybox=\hbox\bgroup\verb}
\newcommand\fverbdo{\egroup\medskip\noindent\fbox{\unhbox\mybox}\ }
\newcommand\fverbit{\egroup\item[\fbox{\unhbox\mybox}]}
\conference{Asymptotic and scattering for degenerate N-solitons in the Hirota equation}
\abstract{We construct all higher order conserved charges from a general two-dimensional zero curvature condition using a Gardner transformation.
Employing two of those charges in the definition of a Hamiltonian allows to view the Hirota equations as an integrable $\mathcal{PT}$-symmetric extension of the 
nonlinear Schr\"{o}dinger equation. 
We construct new degenerate multi-soliton solutions from Hirota's direct method as well as Darboux-Crum transformations based on Jordan states. We
study the properties of these solutions, computing their asymptotic time-dependent displacements and also show that
their scattering process has a distinct characteristic behaviour different from the nondegenerate counterparts allowing only for interactions of absorb-emit type.}

\title{Asymptotic and scattering behaviour for degenerate multi-solitons in
the Hirota equation }
\author{Julia Cen and Andreas Fring \\
Department of Mathematics, City University London,\\
Northampton Square, London EC1V 0HB, UK \\
E-mail: julia.cen.1@city.ac.uk, a.fring@city.ac.uk}

\input{tcilatex}
\begin{document}

\section{Introduction}

The Hirota equation \cite{hirota1973exact} is a well known integrable higher
order extension of the nonlinear Schr\"{o}dinger equation (NLSE) \cite%
{shabat1972exact}. The physical motivation for such an extension was to
allow for a more precise description for the wave propagation of pulses in
the picosecond regime \cite{mollenauer1980}, as the NLSE failed to provide
full explanations of some experiments in the high-intensity and short pulse
subpicosecond regime \cite{mitschke1986,gordon1986}. Mathematically this
equation is of special interest as it constitutes one of the very few
examples for which such type of extensions preserve the integrability. Other
known examples of integrable extensions of the NLSE are the NLSE of type I 
\cite{anderson1983}, the NLSE of type II \cite{chen1979}, the
Hirota-modified Korteweg-de Vries equation \cite{anco2011interaction} and
the Sasa-Satsuma equation \cite{sasa1991}.

Here we view the full Hirota equation as $\mathcal{PT}$-symmetrically
extended version of NLSE. This simple symmetry property together with the
integrability of the system allows for an easy explanation of why the
physical quantities associated to the model are real despite the fact that
they are computed from complex solutions. We extend here our previous
argumentation \cite{CenFring,cen2016time} applied only to the energy of the
system to all charges.

Our main focus in this manuscript is the continuation of the study of
multi-soliton solutions that have the same speed parameters \cite%
{CorreaFring,CCFsineG} leading to identical energies of the their
one-soliton constituents in their multi-soliton solutions, hence they were
referred to as degenerate multi-soliton solutions. Such type of solutions
have been found previously for the NLSE in the context of the inverse
scattering method \cite{multipole,schiebold2017}, where they were referred
to as multiple pole solutions. This terminology is somewhat misleading as
the poles are not actually in the solutions of the NLSE but in the kernels
of the Gel'fand-Levitan-Marchenko equations, that is a specific quantity
within the context of the inverse scattering method. In our previous
analysis for the Korteweg de-Vries equation \cite{CorreaFring} and the
sine-Gordon equation \cite{CCFsineG} we showed how to derive these type of
solutions in a more transparent way by employing Hirota's direct method,
Darboux-Crum transformations or recursive equations derived from B\"{a}%
cklund transformations. Here we follow a similar approach for the Hirota
equation in the construction of the degenerate multi-soliton solutions.

Our manuscript is organized as follows: In section 2 we employ a Gardner
transformation \cite{kupershmidt1981nature} to construct all charges related
to a particular two dimensional form of the AKNS-equation. We use two of
these charges to construct a Hamiltonian that allows to view the Hirota
equations as a $\mathcal{PT}$-symmetrically extended version of NLSE. In
section 3 we construct multi-soliton solutions by means of Hirota's direct
method and Darboux-Crum transformation based on Jordan states. In section 4
we study the properties of these solutions. In particular, we compute closed
expressions for all higher order charges resulting from concrete
multi-solitons solutions, we compute the time-dependent displacements for
the one-soliton constituents in the multi-soliton solutions. We also show
that unlike as in the degenerate case the scattering process for the
degenerate solution only allows for an absorb-emit process. Our conclusions
are stated in section 5.

\section{The Hirota equation as a $\mathcal{PT}$-symmetrically extended NLSE}

We consider here the full Hirota equation \cite{hirota1973exact} in the form%
\begin{equation}
\!iq_{t}\!=\!-\alpha \left[ q_{xx}+2\left\vert q\right\vert ^{2}q\right]
\!-i\beta \!\left[ q_{xxx}+6\left\vert q\right\vert ^{2}q_{x}\right] ,
\label{NLHE}
\end{equation}%
with real constants $\alpha $, $\beta $ and complex valued field $q=q(x,t)$
depending on the position $x$ and time $t$. This equation is known to unify
the modified Korteweg-de Vries (mKdV) equation and the NLSE equation, which
are obtained from it in the limits $\alpha \rightarrow 0$ and $\beta
\rightarrow 0$, respectively. The equation (\ref{NLHE}) is symmetric with
respect to the anti-linear map $\mathcal{PT}:x\rightarrow -x$, $t\rightarrow
-t$, $i\rightarrow -i$, $q\rightarrow q$. The term proportional to $\beta $
can be viewed as a $\mathcal{PT}$-symmetric extension of the NLSE. Evidently
there exist many such choices and so we briefly explain the origin of the
particular form of this extension term that guaranteed the integrability of
the model by constructing the Hamiltonian that corresponds to (\ref{NLHE})
and also all higher order conserved quantities.

We recall that equivalently to the AKNS equation \cite{AKNS}, the Hirota
equation results as a compatibility equation for the two linear first order
differential equations 
\begin{equation}
\Psi _{t}=V\Psi \text{\qquad and\qquad\ }\Psi _{x}=U\Psi ,  \label{aux}
\end{equation}%
with auxiliary function $\Psi $ and operators $U$, $V$ of the form 
\begin{equation}
\Psi =\left( 
\begin{array}{c}
\varphi \\ 
\phi%
\end{array}%
\right) ,\qquad U=\left( 
\begin{array}{cc}
-i\lambda & q \\ 
r & i\lambda%
\end{array}%
\right) ,\qquad V=\left( 
\begin{array}{cc}
A & B \\ 
C & -A%
\end{array}%
\right) ,  \label{aux2}
\end{equation}%
with complex valued scalar functions $r$, $q$, $A$, $B$ and $C$. From this
starting point the conserved quantities for this system are easily derived
from an analogue to the Gardner transform for the KdV field \cite%
{miura1968korteweg,Miura,kupershmidt1981nature,cen2016time}. Defining two
new complex valued fields $T(x,t)$ and $\chi (x,t)$ in terms of the
components of the auxiliary field $\Psi $ one trivially obtains a local
conservation law 
\begin{equation}
T:=\frac{\varphi _{x}}{\varphi },\qquad \chi :=-\frac{\varphi _{t}}{\varphi }%
,\quad \Rightarrow ~~~T_{t}+\chi _{x}=0.  \label{TXI}
\end{equation}%
From the two first rows in the equations (\ref{aux}) we then derive%
\begin{equation}
T=q\frac{\phi }{\varphi }-i\lambda \text{,~~~~~~\ \ ~}\chi =-A-B\frac{\phi }{%
\varphi },
\end{equation}%
so that the local conservation law in (\ref{TXI}) is expressed in terms of
the as yet unknown quantities $A$, $B$ and $T$%
\begin{equation}
T_{t}-\left( A+i\lambda B+\frac{B}{q}T\right) _{x}=0.  \label{LC}
\end{equation}%
The missing function $T$ is then determined by the Ricatti equation 
\begin{equation}
T_{x}=i\lambda \frac{q_{x}}{q}+rq-\lambda ^{2}+\frac{q_{x}}{q}T-T^{2},
\label{Ric}
\end{equation}%
which in turn is obtained by differentiating $T$ in (\ref{TXI}) with respect
to $x$. The Gardner transformation \cite%
{miura1968korteweg,Miura,kupershmidt1981nature,cen2016time} consists now of
expanding $T$ in terms of $\lambda $ and a new field $w$ as $T=-i\lambda
\lbrack 1-w/(2\lambda ^{2})]$. This choice is motivated by balancing the
first with the fourth and the third and the fifth term when $\lambda
\rightarrow \infty $. The factor on the field $w$ is just convenience that
renders the following calculations in a simple form. Substituting this
expression for $T$ into the Ricatti equation (\ref{Ric}) with a further
choice $\lambda =i/(2\varepsilon )$, made once more for convenience, yields%
\begin{equation}
w+\varepsilon \left( w_{x}-\frac{q_{x}}{q}w\right) +\varepsilon
^{2}w^{2}-rq=0.  \label{ww}
\end{equation}%
Up to this point our discussion is entirely generic and the functions $%
r(x,t) $ and $q(x,t)$ can in principle be any function. Fixing their mutual
relation now to $r(x,t)=-q^{\ast }(x,t)$ and expanding the new auxiliary
density field as%
\begin{equation}
w(x,t)=\dsum\limits_{n=0}^{\infty }\varepsilon ^{n}w_{n}(x,t),
\end{equation}%
we can solve (\ref{ww}) for the functions $w_{n}$ in a recursive manner
order by order in $\varepsilon $. Iterating these solutions yields%
\begin{equation}
w_{n}=\frac{q_{x}}{q}w_{n-1}-(w_{n-1})_{x}-\sum%
\limits_{k=0}^{n-2}w_{k}w_{n-k-2},\qquad \text{for }n\geq 1.  \label{wn}
\end{equation}%
We compute the first expressions to%
\begin{eqnarray}
w_{0} &=&-\left\vert q\right\vert ^{2},  \label{w0} \\
w_{1} &=&\frac{1}{2}\left\vert q\right\vert _{x}^{2}+\frac{1}{2}\left(
qq_{x}^{\ast }-q^{\ast }q_{x}\right) , \\
w_{2} &=&\left\vert q_{x}\right\vert ^{2}-\left\vert q\right\vert ^{4}-\frac{%
1}{2}\left( qq_{x}^{\ast }+q^{\ast }q_{x}\right) _{x}+\frac{1}{2}\left(
q^{\ast }q_{xx}-qq_{xx}^{\ast }\right) ,  \label{w2} \\
w_{3} &=&\left[ \frac{5}{4}\left\vert q\right\vert ^{4}+\frac{1}{2}\left(
qq_{xx}^{\ast }+q^{\ast }q_{xx}-\left\vert q_{x}\right\vert ^{2}\right) %
\right] _{x}+\frac{1}{2}\left( qq_{xx}^{\ast }-q^{\ast }q_{xx}\right) _{x}
\label{w3} \\
&&+\frac{1}{2}\left( 3q\left\vert q\right\vert ^{2}q_{x}^{\ast }-3q^{\ast
}\left\vert q\right\vert ^{2}q_{x}+q_{x}^{\ast }q_{xx}-q_{x}q_{xx}^{\ast
}\right) .  \notag
\end{eqnarray}%
When possible we have also extracted terms that can be written as
derivatives, since they become surface terms in the expressions for the
conserved quantities, and also those that give a zero contribution to the
variation. We note that with regard to the aforementioned $\mathcal{PT}$%
-symmetry we have $\mathcal{PT}(w_{n})=(-1)^{n}w_{n}$. Since $T$ is a
density of a local conservation law, also each function $w_{n}$ can be
viewed as a density. We may then define a Hamiltonian density from the two
conserved quantities $w_{2}$ and $w_{3}$ as%
\begin{eqnarray}
\mathcal{H}(q,q_{x},q_{xx}) &\mathcal{=}&\mathcal{\alpha }w_{2}+i\beta w_{3}
\label{HH1} \\
&=&\mathcal{\alpha }\left( \left\vert q_{x}\right\vert ^{2}-\left\vert
q\right\vert ^{4}\right) -i\frac{\beta }{2}\left( q_{x}q_{xx}^{\ast
}-q_{x}^{\ast }q_{xx}\right) -i\frac{3\beta }{4}\left[ \left( q^{\ast
}\right) ^{2}\left( q^{2}\right) _{x}-q^{2}\left( q^{\ast }\right) _{x}^{2}%
\right] ,~~~~~~~  \label{HH2}
\end{eqnarray}%
with some real constants $\alpha $, $\beta $, where we have dropped all
surface terms in (\ref{HH2}) and terms with zero variation, such as the last
one in (\ref{w2}). We also included an $i$ in front of the $w_{3}$-term to
ensure the overall $\mathcal{PT}$-symmetry of $\mathcal{H}$, which prompts
us to view the Hirota equation as a $\mathcal{PT}$-symmetric extension of
the NLSE. This form will ensure the reality of the total energy of the
system, defined by $E(q):=\int\nolimits_{-\infty }^{\infty }$ $\mathcal{H}%
(q,q_{x},q_{xx})dx$ for a particular solution. It is clear from our analysis
that the extension term needs to be of a rather special form as most terms,
even when they respect the $\mathcal{PT}$-symmetry, will destroy the
integrability of the model, see also \cite{fring2013pt} for other models.

It is now easy to verify that equation (\ref{NLHE}) and its conjugate result
from varying the Hamiltonian $H=\int $ $\mathcal{H}dx$%
\begin{equation}
\!~~~iq_{t}\!=\frac{\delta H}{\delta q^{\ast }}=\sum\nolimits_{n=0}^{\infty
}(-1)^{n}\frac{d^{n}}{dx^{n}}\frac{\partial \mathcal{H}}{\partial
q_{nx}^{\ast }},~~\ \ ~\!iq_{t}^{\ast }\!=-\frac{\delta H}{\delta q}%
=\sum\nolimits_{n=0}^{\infty }(-1)^{n}\frac{d^{n}}{dx^{n}}\frac{\partial 
\mathcal{H}}{\partial q_{nx}},
\end{equation}%
with Hamiltonian density (\ref{HH2}). At this point we also determine the
functions%
\begin{eqnarray}
A &=&i\alpha \left\vert q\right\vert ^{2}-2i\alpha \lambda ^{2}+\beta \left(
qq_{x}^{\ast }-q^{\ast }q_{x}-4i\lambda ^{3}+2i\lambda \left\vert
q\right\vert ^{2}\right) , \\
B &=&i\alpha q_{x}+2\alpha \lambda q+\beta \left( 2i\lambda
q_{x}-2q\left\vert q\right\vert ^{2}-q_{xx}+4\lambda ^{2}q\right) , \\
C &=&i\alpha q_{x}^{\ast }-2\alpha \lambda q^{\ast }+\beta \left(
q_{xx}^{\ast }-2q^{\ast }\left\vert q\right\vert ^{2}+2i\lambda q_{x}^{\ast
}-4\lambda ^{2}q^{\ast }\right) ,
\end{eqnarray}%
as solutions to the auxiliary equation (\ref{aux}) up to the Hirota
equation~(\ref{NLHE}). They serve to compute the function $\chi $ occurring
in the local conservation law (\ref{LC}).

\section{Construction of degenerate multi-soliton solutions}

\subsection{Hirota's direct method}

Hirota's direct method is one of the most transparent and straightforward
techniques to find solutions to nonlinear differential equations. We briefly
recall the main principle of this method and utilize it to solve Hirota's
equation (\ref{NLHE}) with a particular focus on how to obtain new
degenerate solutions in this context. Factorizing the complex field in (\ref%
{NLHE}) as $q(x,t)=g(x,t)/f(x,t)$, with $g(x,t)\in \mathbb{C}$, $f(x,t)\in 
\mathbb{R}$, it is well known \cite{hirota1973exact} that one can express
Hirota's equation (\ref{NLHE}) in bilinear form as%
\begin{eqnarray}
iD_{t}g\cdot f+\alpha D_{x}^{2}g\cdot f+i\beta D_{x}^{3}g\cdot f &=&0,
\label{H1} \\
D_{x}^{2}f\cdot f &=&2\left\vert g\right\vert ^{2},  \label{H2}
\end{eqnarray}%
with $D_{x}^{n}$, $D_{t}^{n}$ denoting Hirota derivatives \cite%
{hirota2004direct} defined by an analogue to the Leibniz rule, albeit with
alternating signs,%
\begin{equation}
D_{x}^{n}f\cdot g=\sum\limits_{k=0}^{n}\left( 
\begin{array}{l}
n \\ 
k%
\end{array}%
\right) (-1)^{k}\frac{\partial ^{n-k}}{\partial x^{n-k}}f(x)\frac{\partial
^{k}}{\partial x^{k}}g(x).
\end{equation}%
Exact multi-soliton solutions can be found in a recursive fashion by
terminating the formal power series expansions%
\begin{equation}
f(x,t)=\dsum\limits_{k=0}^{\infty }\varepsilon ^{2k}f_{2k}(x,t),\quad \text{%
and\quad }g(x,t)=\dsum\limits_{k=1}^{\infty }\varepsilon
^{2k-1}g_{2k-1}(x,t),  \label{expfg}
\end{equation}%
at a particular order in $\varepsilon $. The remarkable and well-known
feature of this seemingly perturbative approach is that the solutions
obtained in this manner are exact for any value of the expansion parameter $%
\varepsilon $, when the series are suitably terminated.

\subsubsection{One-soliton solution}

For $\varepsilon =1$ a one-soliton is obtained as%
\begin{equation}
q_{1}^{\mu }(x,t)=\frac{g_{1}^{\mu }(x,t)}{1+f_{2}^{\mu }(x,t)},\quad \ \ 
\text{with~\ }g_{1}^{\mu }(x,t)=\tau _{\mu ,c}\text{\quad and\quad }%
f_{2}^{\mu }(x,t)=\frac{\left\vert \tau _{\mu ,c}\right\vert ^{2}}{(\mu +\mu
^{\ast })^{2}}.  \label{ones}
\end{equation}%
The building blocks are the functions 
\begin{equation}
\tau _{\mu ,c}(x,t):=c\tilde{\tau}_{\mu }(x,t),\qquad \tilde{\tau}_{\mu
}(x,t):=e^{\mu x+\mu ^{2}(i\alpha -\beta \mu )t},
\end{equation}%
involving the complex constants $c$,$\mu \in \mathbb{C}$. More explicitly,
for $c=1$ we have%
\begin{equation}
q_{1}^{\mu }(x,t)=\frac{4\delta ^{2}e^{x(\delta +i\xi )+it(\delta +i\xi
)^{2}(\alpha +i\beta \delta -\beta \xi )}}{4\delta ^{2}+e^{2\delta x-2\delta
t\left[ 2\alpha \xi +\beta \left( \delta ^{2}-3\xi ^{2}\right) \right] }}%
,~~\ \ \ ~\left\vert q_{1}^{\mu }(x,t)\right\vert =\frac{4\delta
^{2}e^{\delta \left[ x-t\left( 2\alpha \xi +\beta \left( \delta ^{2}-3\xi
^{2}\right) \right) \right] }}{4\delta ^{2}+e^{2\delta \left[ x-t\left(
2\alpha \xi +\beta \left( \delta ^{2}-3\xi ^{2}\right) \right) \right] }}.
\label{oneS}
\end{equation}%
with $\mu =\delta +i\xi $, $\delta $,$\xi \in \mathbb{R}$. Defining the real
quantities 
\begin{eqnarray}
A(x,t) &:&=x\xi +t\left[ \alpha (\delta ^{2}-\xi ^{2})+\beta \xi (\xi
^{2}-3\delta ^{2})\right] ,~~~  \label{Ax} \\
x_{\pm }^{\delta ,\xi } &:&=t\left[ 2\alpha \xi +\beta (\delta ^{2}-3\xi
^{2})\right] \pm \frac{1}{\delta }\ln (2\delta ),
\end{eqnarray}%
we compute the maximum of the modulus for the one-soliton solution to 
\begin{equation}
q_{1}^{\mu }(x+x_{+}^{\delta ,\xi },t)=\delta \func{sech}(x\delta
)e^{iA(x+x_{+}^{\delta ,\xi },t)}~~~\ ~~\ \ \left\vert q_{1}^{\mu
}(x_{+}^{\delta ,\xi },t)\right\vert =\delta .  \label{q1}
\end{equation}%
Thus while the real and imaginary parts of the one-soliton solution exhibit
a breather like behaviour, the modulus is a proper solitary wave with a
stable maximum at $\delta $. The solution $q_{1}^{\mu }$ becomes static in
the limit to the NLSE $\beta \rightarrow 0$ for real $\mu $, i.e. $\xi =0$,
and also in the limit to the mKdV equation $\alpha \rightarrow 0$ when $%
\delta ^{2}=3\xi ^{2}$.

\subsubsection{Nondegenerate and degenerate two-soliton solution}

At the next order in $\varepsilon $ of the expansions (\ref{expfg}) we
construct a general nondegenerate two-soliton solution as%
\begin{equation}
q_{2}^{\mu ,\nu }(x,t)=\frac{g_{1}^{\mu ,\nu }(x,t)+g_{3}^{\mu ,\nu }(x,t)}{%
1+f_{2}^{\mu ,\nu }(x,t)+f_{4}^{\mu ,\nu }(x,t)},\quad \ \   \label{twos}
\end{equation}%
with functions%
\begin{eqnarray}
g_{1}^{\mu ,\nu } &=&\tau _{\mu ,c}+\tau _{\nu ,\tilde{c}},  \label{g1} \\
g_{3}^{\mu ,\nu } &=&\frac{\left( \mu -\nu \right) ^{2}}{\left( \mu +\mu
^{\ast }\right) ^{2}\left( \nu +\mu ^{\ast }\right) ^{2}}\tau _{\nu ,\tilde{c%
}}\left\vert \tau _{\mu ,c}\right\vert ^{2}+\frac{\left( \mu -\nu \right)
^{2}}{\left( \mu +\nu ^{\ast }\right) ^{2}\left( \nu +\nu ^{\ast }\right)
^{2}}\tau _{\mu ,c}\left\vert \tau _{\nu ,\tilde{c}}\right\vert ^{2}, \\
f_{2}^{\mu ,\nu } &=&\frac{\left\vert \tau _{\mu ,c}\right\vert ^{2}}{\left(
\mu +\mu ^{\ast }\right) ^{2}}+\frac{\tau _{\nu ,\delta }\tau _{\mu
,c}^{\ast }}{\left( \nu +\mu ^{\ast }\right) ^{2}}+\frac{\tau _{\mu ,c}\tau
_{\nu ,\tilde{c}}^{\ast }}{\left( \mu +\nu ^{\ast }\right) ^{2}}+\frac{%
\left\vert \tau _{\nu ,\tilde{c}}\right\vert ^{2}}{\left( \nu +\nu ^{\ast
}\right) ^{2}}, \\
f_{4}^{\mu ,\nu } &=&\frac{\left( \mu -\nu \right) ^{2}\left( \mu ^{\ast
}-\nu ^{\ast }\right) ^{2}}{\left( \mu +\mu ^{\ast }\right) ^{2}\left( \nu
+\mu ^{\ast }\right) ^{2}\left( \mu +\nu ^{\ast }\right) ^{2}\left( \nu +\nu
^{\ast }\right) ^{2}}\left\vert \tau _{\mu ,c}\right\vert ^{2}\left\vert
\tau _{\nu ,\tilde{c}}\right\vert ^{2}.  \label{f4}
\end{eqnarray}%
We have set here also $\varepsilon =1$. As was noted previously in \cite%
{CorreaFring,cen2016time,CCFsineG} the limit $\mu \rightarrow \nu $ to the
degenerate case can not be carried out trivially for generic values of the
constants $c$, $\tilde{c}$. However, we find that for the specific choice%
\begin{equation}
c=\frac{\left( \mu +\mu ^{\ast }\right) \left( \mu +\nu ^{\ast }\right) }{%
\left( \mu -\nu \right) },~~~~~\tilde{c}=-\frac{\left( \nu +\nu ^{\ast
}\right) \left( \nu +\mu ^{\ast }\right) }{\left( \mu -\nu \right) },
\label{cc}
\end{equation}%
the limit is nonvanishing for all functions in (\ref{g1})-(\ref{f4}). This
choice is not unique, but the form of the denominators is essential to
guarantee the limit to be nontrivial. With $c$ and $\tilde{c}$ as in (\ref%
{cc}) the limit $\mu \rightarrow \nu $ leads to the new degenerate
two-soliton solution%
\begin{equation}
q_{2}^{\mu ,\mu }(x,t)=\frac{\left( \mu +\mu ^{\ast }\right) \tilde{\tau}%
_{\mu }\left[ (2+\hat{\tau}_{\mu })+(2-\hat{\tau}_{\mu })\left\vert \tilde{%
\tau}_{\mu }\right\vert ^{2}\right] }{1+(2+\left\vert \hat{\tau}_{\mu
}\right\vert ^{2})\left\vert \tilde{\tau}_{\mu }\right\vert ^{2}+\left\vert 
\tilde{\tau}_{\mu }\right\vert ^{4}},  \label{dtwos}
\end{equation}%
where we introduced the function%
\begin{equation}
\hat{\tau}_{\mu }(x,t):=x+\mu t(2i\alpha -3\beta \mu )\left( \mu +\mu ^{\ast
}\right) .
\end{equation}%
We observe the two different timescales in this solution entering through
the functions $\hat{\tau}$ and $\tilde{\tau}$, in a linear and exponential
manner, respectively, which is a typical feature of degenerate solutions.

\subsection{Darboux-Crum transformations}

It is well known that the AKNS equation \cite{AKNS} for many integrable
systems can be converted into an eigenvalue equation involving a Hamiltonian
of Dirac type. In our case we can read the second equation in (\ref{aux}) as 
$H\psi =-\lambda \psi $ with $H=-i\sigma _{3}\partial _{x}+V$, $\sigma _{3}$
denoting a standard Pauli matrix, $V_{12}=V_{21}^{\ast }=iq$ and $%
V_{11}=V_{22}=0$. Taking $H=H_{0}$, Darboux-Crum transformations \cite%
{darboux,crum,matveevdarboux,correahidden,correa2008self,mateos2017perfectly}
for Dirac Hamiltonians \cite{nieto2003,correa2017} consist of iterating the
equations 
\begin{equation}
L_{n}H_{n-1}=H_{n}L_{n}
\end{equation}%
with the help of some intertwining operators $L_{n}$ that we do not specify
here any further. The new Hamiltonians $H_{n}$ satisfy the equations $%
H_{n}\Psi _{n}=-\mu _{n}\Psi _{n}$. Generalizing also the first equation in (%
\ref{aux}) by setting $\Psi \rightarrow \Psi _{n}$, the component equations
become%
\begin{equation}
\begin{array}{ll}
\left( \varphi _{2j-1}\right) _{x}=\mu _{j}\varphi _{2j-1}, & \left( \phi
_{2j-1}\right) _{x}=-\mu _{j}\phi _{2j-1}, \\ 
\left( \varphi _{2j-1}\right) _{t}=2\mu _{j}^{2}(i\alpha -2\beta \mu
_{j})\varphi _{2j-1},~~~\  & \left( \phi _{2j-1}\right) _{t}=-2\mu
_{j}^{2}(i\alpha -2\beta \mu _{j})\phi _{2j-1}, \\ 
\varphi _{2j}=-\phi _{2j-1}^{\ast }, & \phi _{2j}=\varphi _{2j-1}^{\ast },%
\end{array}%
~~~\text{for }j=1,\ldots ,n,  \label{xx}
\end{equation}%
as explained in more detail in \cite{CenFringHir}. The solutions to (\ref{xx}%
) 
\begin{equation}
\varphi _{2j-1}=c_{2j-1}e^{\mu _{j}x+2t\mu _{j}^{2}(i\alpha -2\beta \mu
_{j})}=\phi _{2j}^{\ast },~~~\phi _{2j-1}=\tilde{c}_{2j-1}e^{-\mu
_{j}x-2t\mu _{j}^{2}(i\alpha -2\beta \mu _{j})}=-\varphi _{2j}^{\ast },
\label{solxx}
\end{equation}%
are the basic building blocks for the construction of a $n$-soliton
solution. They can be expressed in a very compact form as 
\begin{equation}
q_{n}=2\frac{\det D_{n}}{\det W_{n}}\,,  \label{genrqn}
\end{equation}%
with $W_{n}$ and $D_{n}$ denoting $2n\times 2n$-matrices. The matrix $W_{n}$
consists of $n$ columns containing $\varphi _{i}$ and its derivatives $%
\varphi _{i}^{(n)}:=\partial ^{n}\varphi _{i}/\partial x^{n}$ for $i=1,...,2n
$ and $n$ columns containing $\phi _{i}$ and its derivatives $\phi
_{i}^{(n)}:=\partial ^{n}\phi _{i}/\partial x^{n}$ 
\begin{equation}
W_{n}=\left( 
\begin{array}{cccccccc}
\varphi _{1}^{(n-1)} & \varphi _{1}^{(n-2)} & \ldots  & \varphi _{1} & \phi
_{1}^{(n-1)} & \ldots  & \phi _{1}^{\,\prime } & \phi _{1} \\ 
\varphi _{2}^{(n-1)} & \varphi _{2}^{(n-2)} & \ldots  & \varphi _{2} & \phi
_{2}^{(n-1)} & \ldots  & \phi _{2}^{\,\prime } & \phi _{2} \\ 
\vdots  & \vdots  & \ddots  & \vdots  & \vdots  & \ddots  & \vdots  & \vdots 
\\ 
\varphi _{2n}^{(n-1)} & \varphi _{2n}^{(n-2)} & \ldots  & \varphi _{2n} & 
\phi _{2n}^{(n-1)} & \ldots  & \phi _{2n}^{\,\prime } & \phi _{2n}%
\end{array}%
\right) .  \label{Wm}
\end{equation}%
The matrix $D_{n}$ is made up of $n-1$ columns containing $\varphi _{i}$ and
its derivatives and $n+1$ columns containing $\phi _{i}$ and its derivatives 
\begin{equation}
D_{n}=\left( 
\begin{array}{cccccccc}
\phi _{1}^{(n-2)} & \phi _{1}^{(n-3)} & \ldots  & \phi _{1} & \varphi
_{1}^{(n)} & \ldots  & \varphi _{1}^{\,\prime } & \varphi _{1} \\ 
\phi _{2}^{(n-2)} & \phi _{2}^{(n-3)} & \ldots  & \phi _{2} & \varphi
_{2}^{(n)} & \ldots  & \varphi _{2}^{\,\prime } & \varphi _{2} \\ 
\vdots  & \vdots  & \ddots  & \vdots  & \vdots  & \ddots  & \vdots  & \vdots 
\\ 
\phi _{2n}^{(n-2)} & \phi _{2n}^{(n-3)} & \ldots  & \phi _{2n} & \varphi
_{2n}^{(n)} & \ldots  & \varphi _{2n}^{\,\prime } & \varphi _{2n}%
\end{array}%
\right) .  \label{Dm}
\end{equation}%
For specific choices of the constants $c_{i}$ and $\tilde{c}_{i}$ involved,
the solutions computed from (\ref{genrqn}) match exactly with the one and
two-soliton solutions derived from Hirota's direct method. Taking for the
solutions $q_{1}$ in (\ref{genrqn}) the constants as $\mu _{1}=(\delta +i\xi
)/2$, $c_{1}=\tilde{c}_{1}=-(2\delta )^{-1}$, we obtain (\ref{oneS}) and
taking $c_{1}=\tilde{c}_{1}=(\mu _{1}-\mu _{2})^{-1}$, $c_{2}=\tilde{c}_{2}=1
$ in the solution $q_{2}$ in (\ref{genrqn}), we get (\ref{twos}) with (\ref%
{cc}) and the identification $\mu _{1}=\mu $, $\mu _{2}=\nu $.

The degenerate solutions can be obtained in principle by taking the limits $%
\mu _{1}\rightarrow \mu _{2}\rightarrow \ldots \rightarrow \mu
_{n}\rightarrow \mu $, which, however, only leads to nontrivial solutions
for very specific choices of the constants $c_{i}$ and $\tilde{c}_{i}$. This
is to be expected given the discussion in the precious section. Here we will
not specify those constants, but follow a slightly different approach. As
pointed out in \cite{CenFringHir}, the nontrivial multi-soliton solutions
can be obtained in an alternative and easier fashion in a closed compact
form by replacing in (\ref{Wm}) and (\ref{Dm}) the standard solutions (\ref%
{solxx}) of (\ref{xx}) with Jordan states%
\begin{eqnarray}
\varphi _{2j-1} &\rightarrow &\partial _{\mu }^{j-1}\varphi _{1}\text{,~~~}%
\phi _{2j-1}\rightarrow \partial _{\mu }^{j-1}\phi _{1}\text{,~~~} \\
\varphi _{2j} &\rightarrow &\partial _{\mu }^{j-1}\varphi _{2}\text{,~~~}%
\phi _{2j}\rightarrow \partial _{\mu }^{j-1}\phi _{2}\text{,~~~ }
\end{eqnarray}%
for $j=1,\ldots ,n$. These states are essentially solutions to the
eigenvalue equation for powers of the Hamitonian operator, see e.g.\ \cite%
{CorreaFring} for more details. Explicitly, the first examples for the
matrices $\tilde{D}_{n}$ and $\tilde{W}_{n}$ related to the degenerate
solutions are%
\begin{equation}
\tilde{D}_{1}=\left( 
\begin{array}{cc}
\varphi _{1}^{\prime } & \varphi _{1} \\ 
\varphi _{2}^{\prime } & \varphi _{2}%
\end{array}%
\right) ,~~~\tilde{W}_{1}=\left( 
\begin{array}{cc}
\varphi _{1} & \phi _{1} \\ 
\varphi _{2} & \phi _{2}%
\end{array}%
\right) ,
\end{equation}%
\begin{equation}
\tilde{D}_{2}=\left( 
\begin{array}{cccc}
\phi _{1} & \varphi _{1}^{\prime \prime } & \varphi _{1}^{\prime } & \varphi
_{1} \\ 
\phi _{2} & \varphi _{2}^{\prime \prime } & \varphi _{2}^{\prime } & \varphi
_{2} \\ 
\phi _{1}^{\prime } & \varphi _{1}^{\prime \prime \prime } & \varphi
_{1}^{\prime \prime } & \varphi _{1}^{\prime } \\ 
\phi _{2}^{\prime } & \varphi _{2}^{\prime \prime \prime } & \varphi
_{2}^{\prime \prime } & \varphi _{2}^{\prime }%
\end{array}%
\right) ,~~~\tilde{W}_{2}=\left( 
\begin{array}{cccc}
\varphi _{1}^{\prime } & \varphi _{1} & \phi _{1}^{\prime } & \phi _{1} \\ 
\varphi _{2}^{\prime } & \varphi _{2} & \phi _{2}^{\prime } & \phi _{2} \\ 
\varphi _{1}^{\prime \prime } & \varphi _{1}^{\prime } & \phi _{1}^{\prime
\prime } & \phi _{1}^{\prime } \\ 
\varphi _{2}^{\prime \prime } & \varphi _{2}^{\prime } & \phi _{2}^{\prime
\prime } & \phi _{2}^{\prime }%
\end{array}%
\right) ,
\end{equation}%
\begin{equation}
\tilde{D}_{3}=\left( 
\begin{array}{cccccc}
\phi _{1}^{\prime } & \phi _{1} & \varphi _{1}^{\prime \prime \prime } & 
\varphi _{1}^{\prime \prime } & \varphi _{1}^{\prime } & \varphi _{1} \\ 
\phi _{2}^{\prime } & \phi _{2} & \varphi _{2}^{\prime \prime \prime } & 
\varphi _{2}^{\prime \prime } & \varphi _{2}^{\prime } & \varphi _{2} \\ 
\phi _{1}^{\prime \prime } & \phi _{1}^{\prime } & \varphi _{1}^{iv} & 
\varphi _{1}^{\prime \prime \prime } & \varphi _{1}^{\prime \prime } & 
\varphi _{1}^{\prime } \\ 
\phi _{2}^{\prime \prime } & \phi _{2}^{\prime } & \varphi _{2}^{iv} & 
\varphi _{2}^{\prime \prime \prime } & \varphi _{2}^{\prime \prime } & 
\varphi _{2}^{\prime } \\ 
\phi _{1}^{\prime \prime \prime } & \phi _{1}^{\prime \prime } & \varphi
_{1}^{v} & \varphi _{1}^{iv} & \varphi _{1}^{\prime \prime \prime } & 
\varphi _{1}^{\prime \prime } \\ 
\phi _{2}^{\prime \prime \prime } & \phi _{2}^{\prime \prime } & \varphi
_{2}^{v} & \varphi _{2}^{iv} & \varphi _{2}^{\prime \prime \prime } & 
\varphi _{2}^{\prime \prime }%
\end{array}%
\right) ,~~~\tilde{W}_{3}=\left( 
\begin{array}{cccccc}
\varphi _{1}^{\prime \prime } & \varphi _{1}^{\prime } & \varphi _{1} & \phi
_{1}^{\prime \prime } & \phi _{1}^{\prime } & \phi _{1} \\ 
\varphi _{2}^{\prime \prime } & \varphi _{2}^{\prime } & \varphi _{2} & \phi
_{2}^{\prime \prime } & \phi _{2}^{\prime } & \phi _{2} \\ 
\varphi _{1}^{\prime \prime \prime } & \varphi _{1}^{\prime \prime } & 
\varphi _{1}^{\prime } & \phi _{1}^{\prime \prime \prime } & \phi
_{1}^{\prime \prime } & \phi _{1}^{\prime } \\ 
\varphi _{2}^{\prime \prime \prime } & \varphi _{2}^{\prime \prime } & 
\varphi _{2}^{\prime } & \phi _{2}^{\prime \prime \prime } & \phi
_{2}^{\prime \prime } & \phi _{2}^{\prime } \\ 
\varphi _{1}^{iv} & \varphi _{1}^{\prime \prime \prime } & \varphi
_{1}^{\prime \prime } & \phi _{1}^{iv} & \phi _{1}^{\prime \prime \prime } & 
\phi _{1}^{\prime \prime } \\ 
\varphi _{2}^{iv} & \varphi _{2}^{\prime \prime \prime } & \varphi
_{2}^{\prime \prime } & \phi _{2}^{iv} & \phi _{2}^{\prime \prime \prime } & 
\phi _{2}^{\prime \prime }%
\end{array}%
\right) ,
\end{equation}%
with $\varphi _{1}=ce^{\mu x+2t\mu ^{2}(i\alpha -2\beta \mu )}=\phi
_{2}^{\ast }$, $\phi _{1}=\tilde{c}_{1}e^{-\mu x-2t\mu ^{2}(i\alpha -2\beta
\mu )}=-\varphi _{2}^{\ast }$. The degenerate $n$-soliton solutions are then
computed as%
\begin{equation}
q_{n}^{n\mu }(x,t)=2\frac{\det \tilde{D}_{n}}{\det \tilde{W}_{n}},
\label{nsol}
\end{equation}%
with only one spectral parameter $\mu $ left.

\section{Properties of degenerate multi-soliton solutions}

\subsection{Real charges from complex solutions}

Let us now verify that all the charges resulting from the densities in (\ref%
{wn}) are real. Defining the charges as the integrals of the charge densities%
\begin{equation}
Q_{n}=\dint\nolimits_{-\infty }^{\infty }w_{n}dx  \label{Qn}
\end{equation}%
we expect from the $\mathcal{PT}$-symmetry behaviour $\mathcal{PT}%
(w_{n})=(-1)^{n}w_{n}$ that $Q_{2n}\in \mathbb{R}$ and $Q_{2n+1}\in i\mathbb{%
R}$. Taking now $q_{1}$ to be in the form (\ref{q1}) and shifting $%
x\rightarrow x+x^{+}$ in (\ref{Qn}), we find from (\ref{wn}) that the only
contribution to the integral comes from the iteration of the first term,
that is%
\begin{equation}
Q_{n}=\dint\nolimits_{-\infty }^{\infty }\left( \frac{q_{x}}{q}\right)
^{n}w_{0}dx.  \label{Qnn}
\end{equation}%
It is clear that the second term in (\ref{wn}), $(w_{n-1})_{x}$, does not
contribute to the integral as it is a surface term. Less obvious is the
cancellation of the remaining terms, which can however be verified easily.
For the one-soliton solution (\ref{q1}) the charges (\ref{Qnn}) become%
\begin{eqnarray}
Q_{n} &=&-\delta ^{2}\dint\nolimits_{-\infty }^{\infty }\left[ i\xi -\delta
\tanh (x\delta )\right] ^{n}\func{sech}^{2}(x\delta )dx \\
&=&-\left\vert \delta \right\vert \dint\nolimits_{-1}^{1}\left( i\xi -\delta
u\right) ^{n}du \\
&=&-\left\vert \delta \right\vert \dsum\nolimits_{k=0}^{n}\frac{n!}{%
(k+1)!(n-k)!}\delta ^{k}(i\xi )^{n-k}\left[ 1+(-1)^{k}\right] .  \label{sum}
\end{eqnarray}%
Since only the terms with even $k$ contribute to the sum in (\ref{sum}), it
is evident from this expression that $Q_{2n}\in \mathbb{R}$ and $Q_{2n+1}\in
i\mathbb{R}$.

Of special interest is the energy of the system resulting from the
Hamiltonian (\ref{HH1}). For the one-soliton solution (\ref{q1}) we obtain 
\begin{equation}
E(q_{1}^{\mu })=\mathcal{\alpha }Q_{2}+i\beta Q_{3}=2\left\vert \delta
\right\vert \left[ \alpha \left( \xi ^{2}-\frac{\delta ^{2}}{3}\right)
+\beta \xi \left( \delta ^{2}-\xi ^{2}\right) \right] .
\end{equation}%
As expected, due to the $\mathcal{PT}$-symmetry the energy is real despite
being computed from a complex field.

The energy of the two-soliton solution (\ref{dtwos}) is computed to 
\begin{equation}
E(q_{2}^{\mu ,\mu })=2E(q_{1}^{\mu }).  \label{e1}
\end{equation}%
The doubling of the energy for the degenerate solution in (\ref{dtwos}) when
compared to the one-soliton solution is of course what we expect from the
fact that the model is integrable and the computation constitutes therefore
an indirect consistency check. We expect (\ref{e1}) to generalize to $%
E(q_{3}^{n\mu })=nE(q_{1}^{\mu })\,$, which we verified numerically for $n=3$
using the solution (\ref{nsol}).

\subsection{Asymptotic behaviour}

Next we compute the asymptotic displacement in the scattering process in a
similar fashion as discussed in more detail in \cite%
{CorreaFring,cen2016time,CCFsineG}. The analysis relies on computing the
asymptotic limits of the multi-soliton solutions and comparing the results
with the tracked one-soliton solution. As a distinct point we track the
maxima of the one-soliton solution (\ref{ones}) within the two-soliton
solution. Similarly as the one-soliton, the real and imaginary parts of the
two-soliton solution depend on the function $A(x,t)$, as defined in (\ref{Ax}%
), occurring in the argument of the $\sin $ and $\cos $ functions. This
makes it is impossible to track a distinct point with constant amplitude.
However, as different values for $A$ only produce an internal oscillation we
can fix $A$ to any constant value without affecting the overall speed.

We start with the calculation for the degenerate two-soliton solution and
illustrate the above behaviour in figure \ref{Fig1} for a concrete choice of 
$A$.

\FIGURE{ \epsfig{file=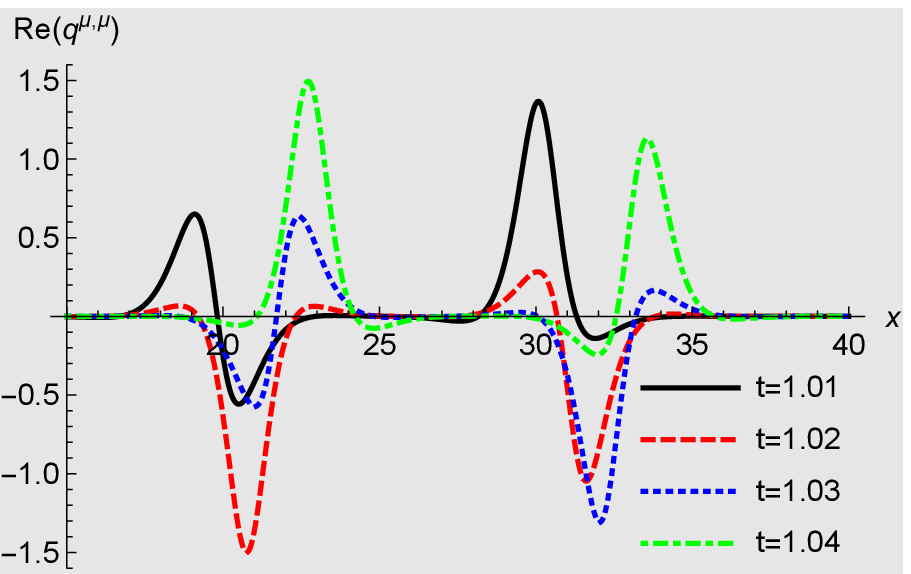,width=7.25cm} \epsfig{file=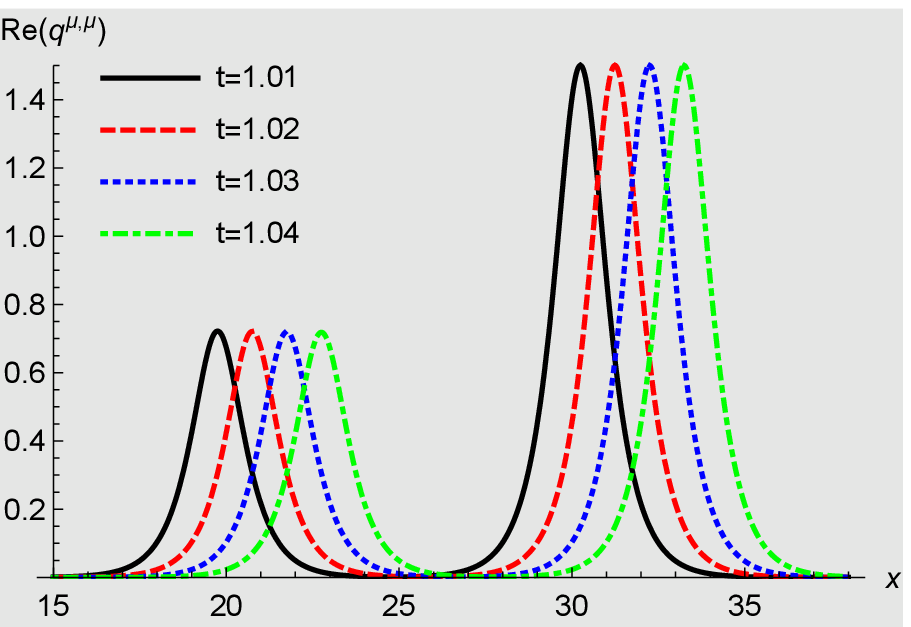,width=7.25cm}
\caption{Real part of the degenerate two-soliton solution (\ref{dtwos}) for the Hirota equations (\ref{NLHE}) at small values of times for $\alpha = 1$, $\beta=2$, 
        $\delta=3/2$, $\xi=1$ with generic $A(x,t)$ in the left panel and fixed $A=\pi/3$ in the right panel.}
        \label{Fig1}}

The functions with constant values of $A$ can be seen as enveloping
functions similar to those employed for the computation of displacements in
breather functions, see e.g. \cite{CCFsineG}. Thus with $A(x,t)=A$ taken to
be constant we calculate the four limits 
\begin{eqnarray*}
\lim_{t\rightarrow \pm \infty }\!q_{2}^{\mu ,\mu }(x_{+}^{\delta ,\xi
}+\Delta (t),t) &=&\pm \frac{\beta \delta ^{2}\cos A-\delta (\alpha -3\beta
\xi )\sin A}{\sqrt{\beta ^{2}\delta ^{2}+(\alpha -3\beta \xi )^{2}}}\pm i%
\frac{\delta (\alpha -3\beta \xi )\cos A+\beta \delta ^{2}\sin A}{\sqrt{%
\beta ^{2}\delta ^{2}+(\alpha -3\beta \xi )^{2}}} \\
\lim_{t\rightarrow \pm \infty }\!q_{2}^{\mu ,\mu }(x_{-}^{\delta ,\xi
}-\Delta (t),t) &=&\mp \frac{\beta \delta ^{2}\cos A+\delta (\alpha -3\beta
\xi )\sin A}{\sqrt{\beta ^{2}\delta ^{2}+(\alpha -3\beta \xi )^{2}}}\pm i%
\frac{\delta (\alpha -3\beta \xi )\cos A-\beta \delta ^{2}\sin A}{\sqrt{%
\beta ^{2}\delta ^{2}+(\alpha -3\beta \xi )^{2}}}
\end{eqnarray*}%
with time-dependent displacement 
\begin{equation}
\Delta (t)=\frac{1}{\delta }\ln \left[ 2\delta \left\vert t\right\vert \sqrt{%
\beta ^{2}\delta ^{2}+(\alpha -3\beta \xi )^{2}}\right] .  \label{Dt}
\end{equation}%
Using the limits form above we obtain the same asymptotic value in all four
cases for the displaced modulus of the two-soliton solution 
\begin{equation}
\lim_{t\rightarrow \pm \infty }\left\vert q_{2}^{\mu ,\mu }(x_{\pm }^{\delta
,\xi }\pm \Delta (t),t)\right\vert =\delta .
\end{equation}%
In the limit to the NLSE, i.e. $\beta \rightarrow 0$, our expression for $%
\Delta (t)$ agrees precisely with the result obtained in \cite{multipole}.

We have here two options to interpret these calculations: As the compound
two-soliton structure is entirely identical in the two limits $t\rightarrow
\pm \infty $ and its individual one-soliton constituents are
indistinguishable we may conclude that there is no overall displacement for
the individual one-soliton constituents. Alternatively we may assume that
the two one-soliton constituents have exchanged their position and thus the
overall time-dependent displacement is $2/\delta \ln (2\delta )+2\Delta (t)$.

For comparison we compute next the displacement for the nondegenerate
two-soliton solution (\ref{twos}) with $c=\tilde{c}=1$ and parameterization $%
\mu =\delta +i\xi $, $\nu =\rho +i\sigma $ where $\delta $,$\xi $,$\rho $,$%
\sigma \in \mathbb{R}$. For definiteness we take $x_{+}^{\delta ,\xi
}>x_{+}^{\rho ,\sigma }$ and calculate the asymptotic limits%
\begin{eqnarray}
\lim_{t\rightarrow +\infty }\left\vert q_{2}^{\mu ,\nu }(x_{+}^{\delta ,\xi
}+\frac{1}{\delta }\tilde{\Delta},t)\right\vert &=&\lim_{t\rightarrow
-\infty }\left\vert q_{2}^{\mu ,\nu }(x_{+}^{\delta ,\xi },t)\right\vert
=\delta , \\
\lim_{t\rightarrow +\infty }\left\vert q_{2}^{\mu ,\nu }(x_{+}^{\rho ,\sigma
},t)\right\vert &=&\lim_{t\rightarrow -\infty }\left\vert q_{2}^{\mu ,\nu
}(x_{+}^{\delta ,\xi }+\frac{1}{\rho }\tilde{\Delta},t)\right\vert =\rho ,
\end{eqnarray}%
with constant%
\begin{equation}
\tilde{\Delta}=\ln \left[ \frac{(\delta +\rho )^{2}+(\xi -\sigma )^{2}}{%
(\delta -\rho )^{2}+(\xi -\sigma )^{2}}\right] .  \label{dtilde}
\end{equation}%
Thus, while the faster one-soliton constituent with amplitude $\delta $ is
advanced by the amount $\tilde{\Delta}/\delta $, the slower one-soliton
constituent with amplitude $\rho $ is regressed by the amount $\tilde{\Delta}%
/\rho $. We compare the two-soliton solution with the two one-soliton
solutions in figure \ref{Fig2}.

\FIGURE{ \epsfig{file=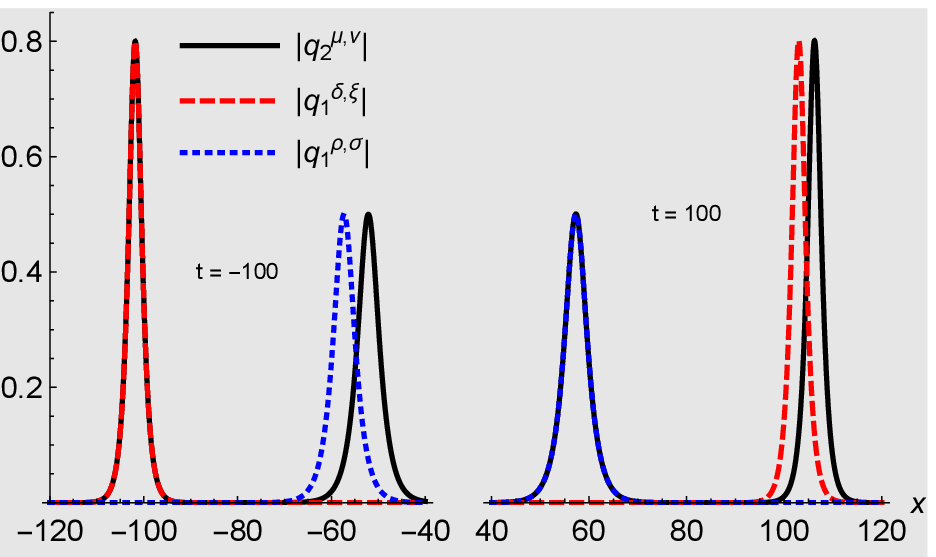,width=7.25cm} \epsfig{file=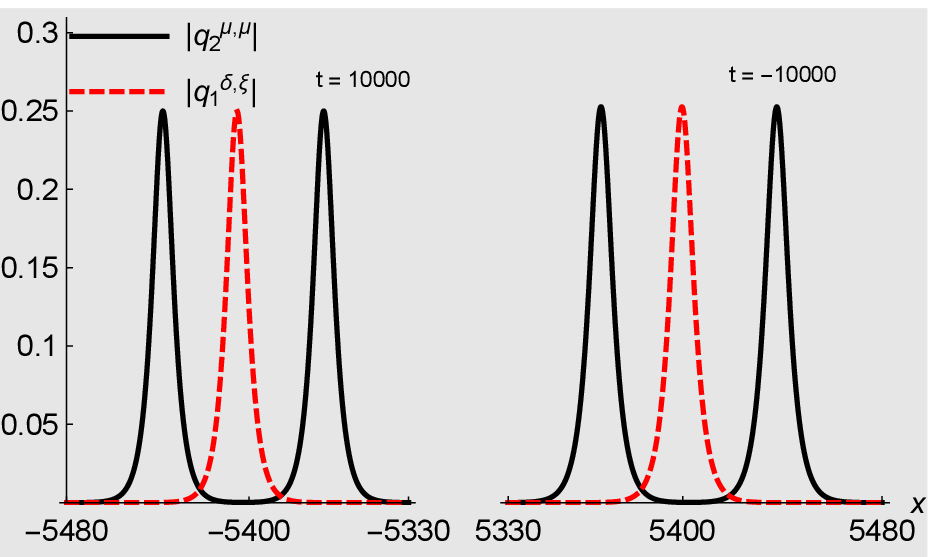,width=7.25cm}
\caption{Nondegenerate two-soliton solution compared to two one-soliton solutions for large values of $|t|$ for $\alpha =1.1$, $\beta =0.9$, $\delta =0.8$, $\xi =0.4$, $\rho =0.5$, $\sigma =0.6$ in the left panel. Degenerate two-soliton solution compared to two one-soliton solutions for large values of $|t|$ for $\alpha =1.5$, $\beta =2.3$, 
$\delta =0.25$, $\xi =0.6$ in the right panel.}
        \label{Fig2}}

We also observe that while the time-dependent displacement $\Delta (t)$ in (%
\ref{Dt}) for the degenerate solution depends explicitly on the parameters $%
\alpha $ and $\beta $, the constant $\tilde{\Delta}$ in (\ref{dtilde}) is
the same for all values of $\alpha $ and $\beta $. In particular it is the
same in the Hirota equation, the NLSE and the mKdV equation. The values for $%
\alpha $ and $\beta $ only enter through $x_{+}^{\rho ,\sigma }$ in the
tracking process.

\subsection{Scattering behaviour}

Besides having a distinct asymptotic behaviour, the degenerate
multi-solitons also display very particular features during the actual
scattering event near $x=t=0$ when compared to the nondegenerate solutions.
For the nondegenerate two-soliton solution three distinct types of
scattering processes at the origin have been identified. Using the
terminology of \cite{anco2011interaction} they are \emph{merge-split}
denoting the process of two solitons merging into one soliton and
subsequently separating while each one-soliton maintains the direction and
momentum of its trajectory, \emph{bounce-exchange} referring to two-solitons
bouncing off each other while exchanging their momenta and \emph{absorb-emit 
}characterizing the process of one soliton absorbing the other at its front
tail and emitting it at its back tail, see figure \ref{ThreeProc}.

\FIGURE{ \epsfig{file=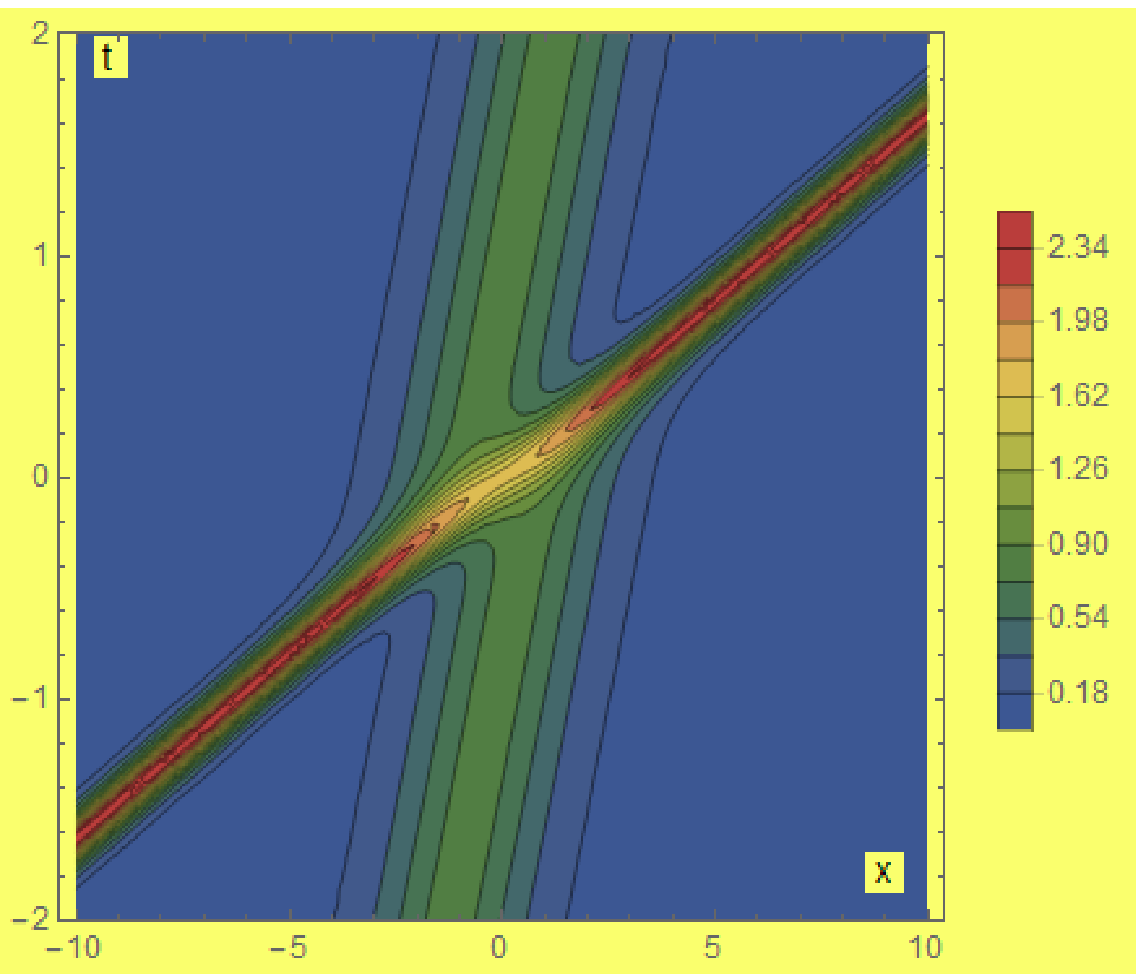,width=5.0cm}\epsfig{file=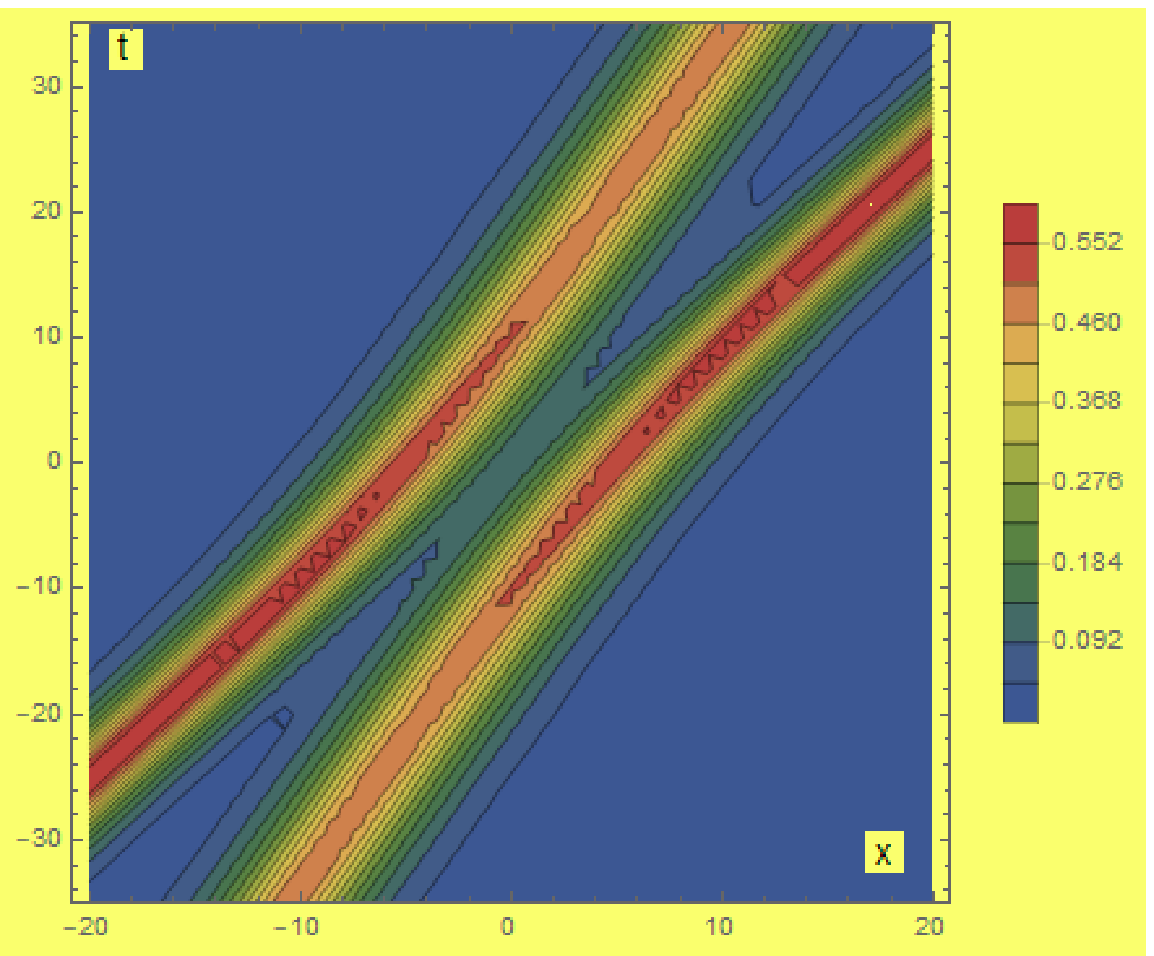,width=5.0cm}\epsfig{file=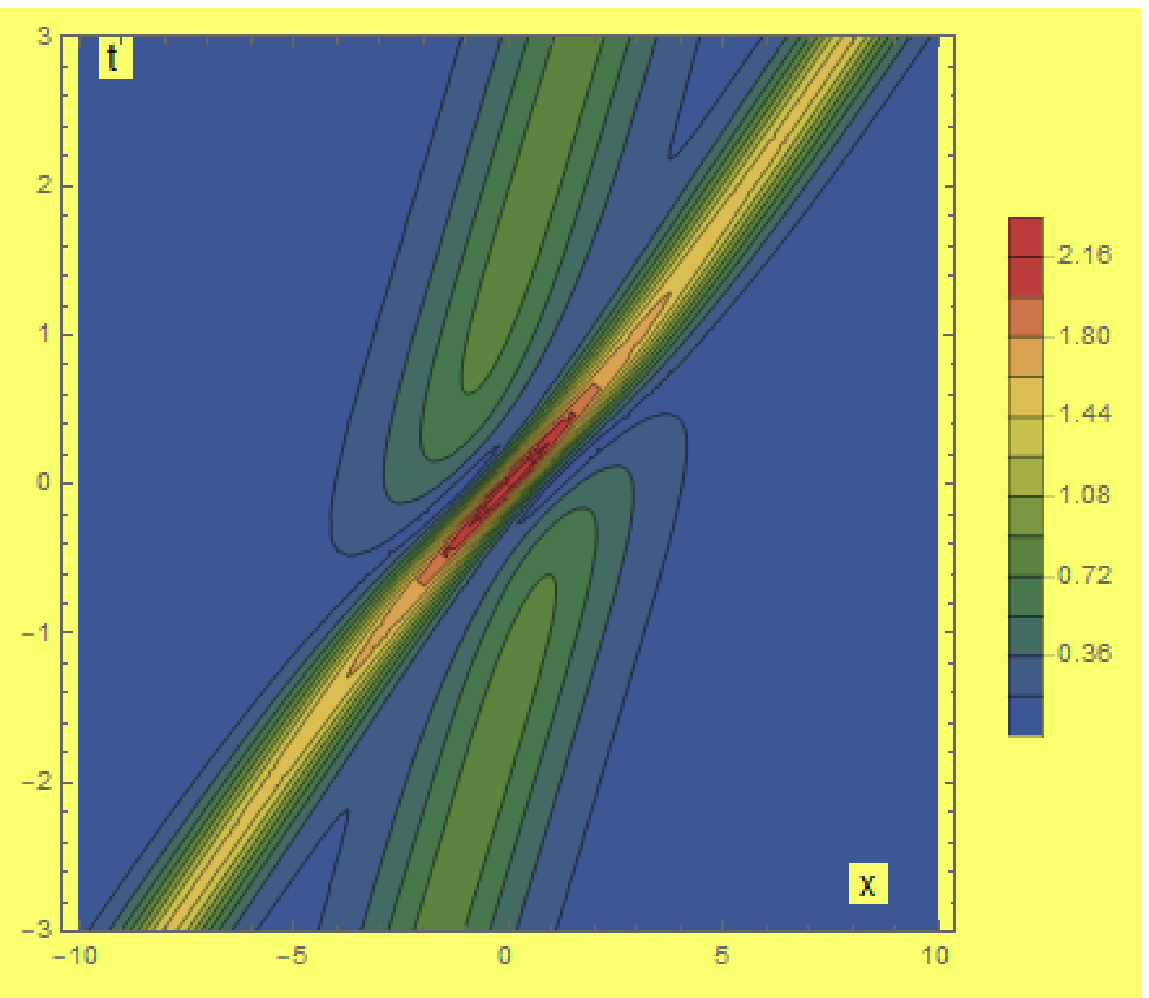,width=5.0cm}
\caption{Different types of nondegenerate two-soliton scattering processes for the solution (\ref{twos}). Left
panel: merge-split scattering with $\alpha =1.1$, $\beta =0.9$, $\rho =2.5$, 
$\xi =0.4$, $\delta =-0.8$, $\sigma =0.6$. Middle panel: bounce-exchange
scattering with $\alpha =1.1$, $\beta =0.9$, $\rho =-0.6$, $\xi =0.1$, $\delta =0.5$, $\sigma =0.2$. Right panel: absorb-emit scattering with $\alpha =1.1$, $\beta =0.9$, $\rho =-1.5$, $\xi =0.4$, $\delta =-0.8$, $\sigma =0.6$.}
        \label{ThreeProc}}

For the degenerate multi-soliton solutions the merge-split and
bounce-exchange scattering is not possible and only the absorb-emit
scattering process occurs as seen in figure \ref{Degscat}.

\FIGURE{ \epsfig{file=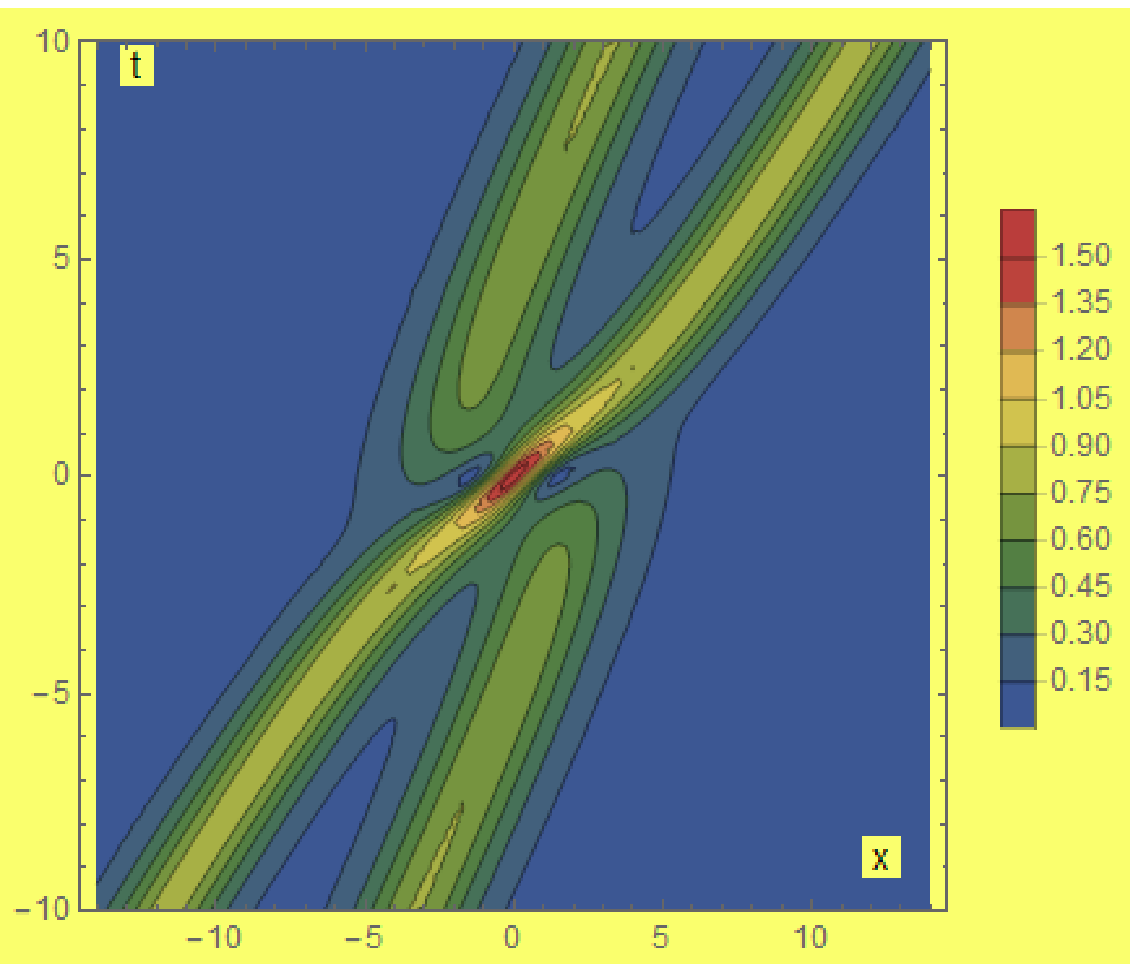,width=7.5cm}\epsfig{file=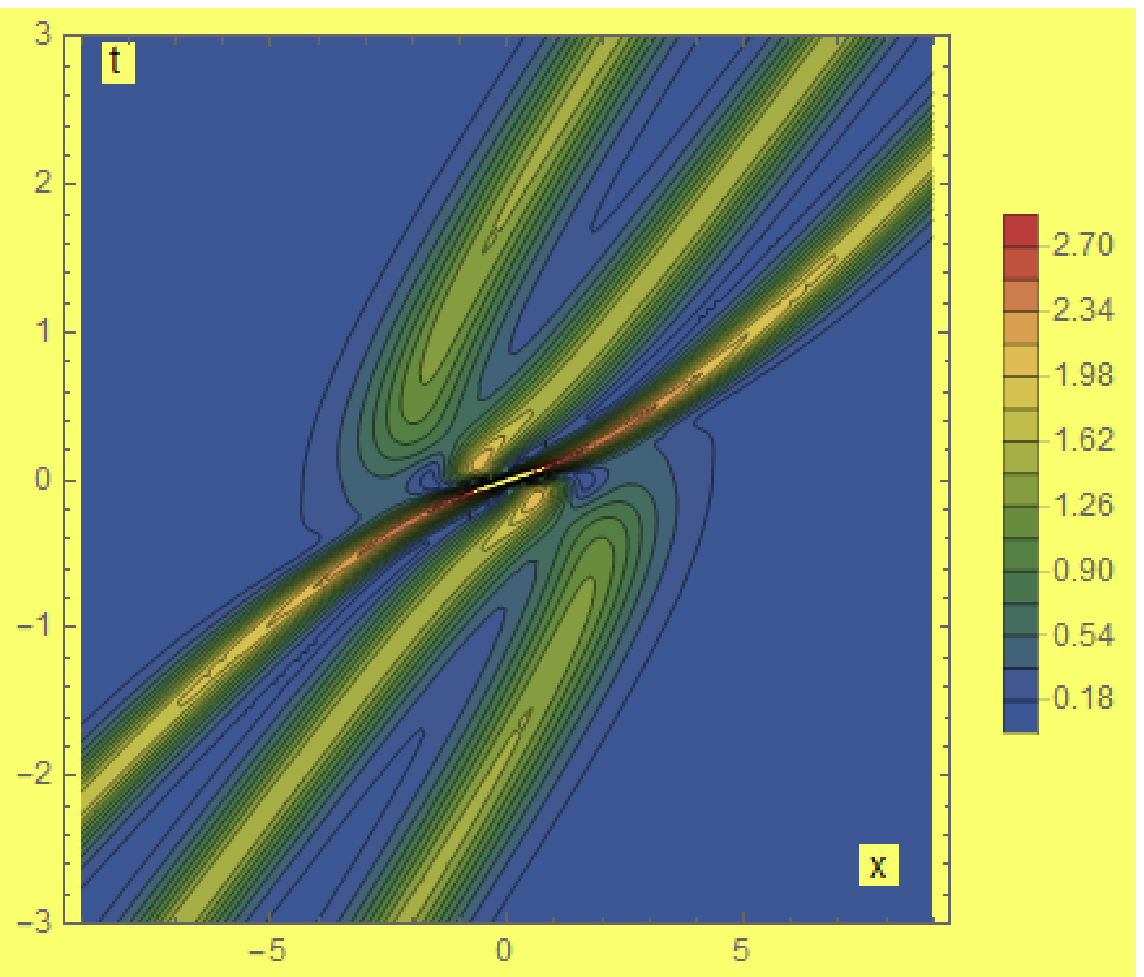,width=7.5cm}
\caption{Absorb-emit scattering processes for degenerate two-solitons (\ref{dtwos}) with $\alpha =1.1$, $\beta =0.9$, $\delta =0.8$, $\xi =0.1$ (left panel) and three-solitons (\ref{nsol}) 
     with $\alpha =1.1$, $\beta =0.9$, $\delta =0.8$, $\xi =0.4$ (right panel). }
        \label{Degscat}}

This feature is easy to understand when considering the behaviour of the
solution at $x=t=0$. As was argued in \cite{anco2011interaction} the
different behaviour can classified by their behaviour as being either convex
downward or concave upward at $x=t=0$ together with occurrence of additional
local maxima. For the degenerate two-soliton solution we find $\left.
\partial \left\vert q_{2}^{\mu ,\mu }(x,t)\right\vert /\partial x\right\vert
_{x=0,t=0}=0$ and $\left. \partial ^{2}\left\vert q_{2}^{\mu ,\mu
}(x,t)\right\vert /\partial x^{2}\right\vert _{x=0,t=0}=-10\left\vert \delta
\right\vert ^{3}$, which means this solution is always concave at $x=t=0$.
In addition, we find that $\func{Re}\left. q_{2}^{\mu ,\mu }(x,t)\right\vert
_{x=0,t=0}$ and $\func{Im}\left. q_{2}^{\mu ,\mu }(x,t)\right\vert _{x=0,t=0}
$ are always concave and convex at $x=t=0$, respectively. Hence, we always
have the emergence of additional local maxima, such that the behaviour must
be of the absorb-emit type. In figure \ref{Degscat} we display this
scattering behaviour for the degenerate two and three-soliton solutions in
which the distinct features of the absorb-emit behaviour are clearly
identifiable.

We observe that the dependence on the parameters $\alpha ~$and $\beta $ of
the degenerate and nondegenerate solution is now reversed when compared to
the asymptotic analysis. While the type of scattering in the nondegenerate
case is highly sensitive with regard to $\alpha ~$and $\beta $, it is
entirely independent of these parameters in the degenerate case.

\section{Conclusions}

We constructed all charges resulting from the AKNS equation (\ref{aux}) and (%
\ref{aux2}) by means of a Gardner transformation. Two of the charges were
used to define a Hamiltonian whose functional variation led to the Hirota
equation. The behaviour of these charges under $\mathcal{PT}$-symmetry
suggests to view the Hirota system as an integrable extended version of
NLSE. This point of view allows for an easy generalization of previous
arguments \cite{CenFring,cen2016time} that guarantee the reality of the
energy to all higher order charges. We computed a closed analytic expression
for all charges involving a particular one-soliton solution.

Explicit multi-soliton solutions from Hirota's direct method as well as the
Darboux-Crum transformations were derived and we showed how to construct
degenerate solutions in both schemes. As observed previously, the
application of Hirota's direct method relies on choosing the arbitrary
constants in the solutions in a very particular way. When using Darboux-Crum
transformations the degenerate solutions are obtained by replacing standard
solutions in the underlying auxiliary eigenvalue problem by Jordan states.

From the asymptotic behaviour of the degenerate two-soliton solution we
computed the new expression for the time-dependent displacement. As the
degenerate one-soliton constituents in the multi-soliton solutions are
asymptotically indistinguishable one can not decide whether the two
one-solitons have actually exchanged their position and therefore the
time-dependent displacement can be interpreted as an advance or delay or
whether the two one-solitons have only approached each other and then
separated again. The analysis of the actual scattering event allows for both
views. It would be very interesting to investigate the statistical behaviour
of a degenerate soliton gas along the lines of, for instance \cite%
{solgas1,solgas2,solgas3,solgas4}, which should certainly exhibit different
characteristics as the underlying statistical distributions would be based
on indistinguishable rather than distinguishable particles.

We showed that degenerate two-solitons may only scatter via an absorb-emit
process, that is by one soliton absorbing the other at its front tail and
subsequently emitting it at the back tail. Since the model is integrable all
multi-particle/soliton scattering processes may be understood as consecutive
two particle/soliton scattering events, so that the two-soliton scattering
behaviour extends to the multi-soliton scattering as we demonstrated.\medskip

\textbf{Acknowledgments:} JC is supported by a City, University of London
Research Fellowship and would like to thank Vincent Caudrelier, Alexander
Mikhailov and Simon Ruijsenaars for discussions and references on the NLSE.
We would also like to thank Francisco Correa for discussions.

\newif\ifabfull\abfulltrue

%%\bibliographystyle{phreport}
%%\bibliography{acompat,Ref}

\end{document}